\documentstyle[preprint,aps,epsf]{revtex}
\global\firstfigfalse
\global\firsttabfalse
\makeatletter
\global\@specialpagefalse
\def\@oddhead{}
\let\@evenhead\@oddhead
\def\@oddfoot{\reset@font\rm \hfill \thepage
{\footnotesize $_{\mbox{(Date:\today)} }$} \hfill}
\let\@evenfoot\@oddfoot
\makeatother

\begin{document}
\draft
\title{DC--Transport in Quantum Wires}
\author{Yuval Oreg and Alexander M. Finkel'stein \cite{Lan}}
\address{Department of Condensed Matter Physics\\
The Weizmann Institute of Science, Rehovot ISRAEL 76100}
\date{\today }
\maketitle

\begin{abstract}
  The influence of electron--electron interaction on two terminal DC
  conductance of one--dimensional quantum wires is studied.  A
  cancelation between the effect of the electron--electron interaction
  on the current and on the external electric field is the reason for
  the universal value, $e^2/2\pi\hbar $ per mode, of the DC
  conductance of a clean wire. The effect of the renormalization of
  the electric field on the DC conductance in the presence of an
  interplay between the electron--electron interaction and backward
  scattering due to an impurity is considered.
\end{abstract}

\pacs{PACS:72.10.-d, 72.15.Nj, 73.40.Hm}

It is well understood now that for non interacting electrons a two
terminal DC conductance, $G$, of a clean quantum wire is $e^2/2\pi
\hbar $ per mode \cite {1DT:Landauer}. Since the electron--electron
(el--el) interaction renormalizes the current, it was generally
accepted after the calculation of Ref.~\cite {1DT:Apel82} that the
conductance should be renormalized to the value $ K_\rho e^2/2\pi
\hbar $, where the parameter $K_\rho $ is related to the
density--density electron interaction ($K_\rho =1$ in the absence of
the interaction). In this paper we explain why despite the fact that
the current is renormalized, the conductance is not. In a clean wire
with an el--el interaction $G=e^2/2\pi \hbar. $   This occurs because
the electric field is also renormalized by the el--el interaction. The
DC conductance of a clean one--dimensional (1D) electron liquid is a
property in which the effect of the el--el interaction on the electric
field cancel out its effect on the current
\cite{QHE:Oreg95,1DT:Kawabata96}.  The question of the renormalization
of the external electric field to the total one demands a special care
in 1D because the only possible electric field is longitudinal .  The
influence of the renormalization of the electric field on the
conductance when there is an interplay between an impurity backward
scattering and the el--el interaction is also discussed.

Recently Tarucha et al. \cite{1DT:Tarucha95} measured the conductance
of a quantum wire formed from Al$_{0.35}$Ga$_{0.65}$As/AlGa modulation
doped heterostructures. Based on the temperature dependence of the
conductance, it was found following the analysis of
Refs.~\cite{1DT:Kane92aFurusaki93Ogata94} that $K_\rho \sim 0.7$.
However, in contrary to the earlier predictions, the
conductance per mode was very close to the universal value $e^2/2\pi \hbar $.
To explain the results of Ref.~\cite{1DT:Tarucha95} the experimental
device was modeled \cite{1DT:Maslov} by a system with two line segments attached to the
central part of the quantum wire. The temperature dependence of the
conductance was controlled by an impurity located in the central part where
the el--el interaction parameter was $K_\rho ^W$, while the
interaction parameter in the attached segments was $K_\rho ^L\neq K_\rho ^W$.
For the length of the segments much larger than the length of the central
part, the quantization of the conductance was determined by $K_\rho ^L$ only,
and was given by $K_\rho ^Le^2/2\pi \hbar $ in this theory. Eventually, it
was assumed that the line segments represent the leads where the electrons
are free, i.e., $K_\rho ^L=1$, and in this way the discrepancy between the
theory and the experiment was settled. On the other hand, it has been argued
by Kawabata \cite{1DT:Kawabata96} and independently by us \cite{QHE:Oreg95}
(in the context of edge states in the quantum Hall regime), that when the
conductance is defined as the response to the total field, rather than to the
external one, the DC conductance of a clean 1D system is not influenced by
the el--el interaction.  The renormalization of the electric
field by the el--el interaction was ignored in Refs.~\cite{1DT:Maslov}, as
well as in earlier publications. In this paper we clarify the role of this
effect.

Let us consider the case of a clean wire when  the density--density electron
interaction exists and may be not homogeneous. First we will show the
universality of the DC conductance in a procedure similar to the one 
elaborated in the context of the edge states in quantum Hall devices \cite
{QHE:Oreg95}. Then, we extend the consideration of Ref.~\cite{1DT:Kane88} to
the case of interacting electrons and show how to calculate the two terminal
conductance in the Kubo formalism.

The DC conductance of a quantum wire connecting two reservoirs (leads) is
given by 
\begin{equation}
G=eI/(\mu _R^1-\mu _L^2)=eI/(\mu _R^2-\mu _L^1),  \label{eq:Con--Lana}
\end{equation}
where $I$ is the current ($I$ does not depend on $x$ in the DC limit), $\mu
_{R(L)}^1$ is the chemical potential of the right (left) moving electrons near
the left reservoir and $\mu _{R(L)}^2$ is the chemical
potential of the right (left) moving electrons near the right reservoir. The
second equality in Eq.~(\ref{eq:Con--Lana}) follows from the fact that in
equilibrium the conductance calculated for electrons or for holes should
give the same result. To find the current $I$ the continuity equation will
be used. In the absence of backward scattering of electrons we can apply the
continuity equation for the left and right moving electrons separately, 
\begin{equation}
J_{R,L}(p)=\frac ip\frac d{dt}e\rho _{R,L}(p)=\frac e{\hbar p}\left[ H,\rho
_{R,L}(p)\right] ,  \label{Eq:Current-operator-1d}
\end{equation}
where $H$ is the Hamiltonian of the system. The density operators $\rho
_{R,L}(p)=\sum_{k\approx \pm k_F}a_{k+p}^{\dagger }a_k$ have the standard 1D
commutation relations \cite{EE1D:Mattis65}: 
\begin{equation}
\renewcommand{\arraystretch}{1.5} 
\begin{array}{c}
\left[ \rho _R(-p),\rho _R(p^{\prime })\right] =\left[ \rho _L(p),\rho
_L(-p^{\prime })\right] =\frac{pL}{2\pi }\delta _{p,p^{\prime }}; \\ 
\left[ \rho _L(p),\rho _R(p^{\prime })\right] =0,
\end{array}
\label{Eq:cr-1d}
\end{equation}
where $L$ is the length of the wire. For operators commuting like that,
performing commutation is equivalent to differentiation, i.e., $\left[
F\left\{ \rho _{R,L}\right\} ,\rho _{R,L}(p)\right] =\pm \frac{pL}{2\pi }{
\partial F\left\{ \rho _{R,L}\right\} }/{\partial \rho _{R,L}(-p)}$. Since
in 1D the Hamiltonian is a functional of the $\rho _{R,L}$ operators, we can
rewrite $J_{R,L}$ in Eq.~(\ref{Eq:Current-operator-1d}) as 
\begin{equation}
\frac 1LJ_{R,L}(p)=\pm \frac e{2\pi \hbar }\frac{\partial H}{\partial \rho
_{R,L}(-p)}.  \label{Eq:current=diffrential-energy-1d}
\end{equation}
On the other hand, by definition, the chemical potentials of the right and
left moving species of electrons are given by 
\begin{equation}
\frac 1L\mu _{R,L}(p)=\frac{\partial H}{\partial \rho _{R,L}(-p)}.
\label{Eq:Chemical-potential-1d}
\end{equation}
Thus, for the total current $J=J_R+J_L$ we obtain that $J(p)=\frac e{2\pi
\hbar }\left( \mu _R(p)-\mu _L(p)\right) $. This result holds for any
momentum $p$ and therefore it can be represented also in space i.e., at any
point $x$ the current $J(x)=\frac e{2\pi \hbar }\left( \mu _R(x)-\mu
_L(x)\right) $. Since in the DC limit the current
 $I=\left\langle J\right\rangle $ does not depend on $x$, 
\begin{equation}
\mu _R^1-\mu _L^1=\mu _R^2-\mu _L^2=(2\pi \hbar /e)I.  \label{eq:Con--Lanb}
\end{equation}
Finally, the combination of Eqs.~(\ref{eq:Con--Lana}) and (\ref{eq:Con--Lanb})
 leads to $\mu _L^1=\mu _L^2=\mu _L,\;\mu _R^1=\mu
_R^2=\mu _R$ in the DC limit, and correspondingly 
\begin{equation}
G=e^2/2\pi \hbar .  \label{eq:Con--Lanc}
\end{equation}
Note, that separately each of $J_R,J_L,\mu _R$ and $\mu _L$ is influenced by
the el--el interaction, while in the particular ratio defining the
conductance the renormalization of the chemical potential difference cancel
out the renormalization of the current.

The above treatment is in the spirit of Landauer's approach \cite{1DT:Landauer}.
 Now we consider the conductance of the 1D
electron gas using the Kubo formalism. In a two terminal measurement the
electrons accelerated by the total electric field inside the wire dissipate
their energy in the reservoirs. The total electric field $E^{tot}(x)$ is
built from the external field and the induced one. Since the electric
field vanishes inside the reservoirs, the DC conductance of the two terminal
device is given by 
\begin{equation}
G=\displaystyle{\int_0^LI(x)E^{tot}(x)dx}\left/ \displaystyle{\left(
\int_0^LE^{tot}(x)dx\right) ^2}\right. .  \label{eq:def-cond}
\end{equation}
Let us define a tensor $\sigma (x,x^{\prime })$, such that 
\begin{equation}
I(x)=\int_0^L\sigma (x,x^{\prime })E^{tot}(x^{\prime })dx^{\prime }.
\label{eq:sigma-def}
\end{equation}
It follows from the Kubo formula that $\sigma (x,x^{\prime })$ is a
divergenceless tensor in the DC limit, i.e., $d\sigma (x,x^{\prime })/dx=0$,
see appendix A of Ref.~\cite{1DT:Kane88}. This property of $\sigma
(x,x^{\prime })$ together with Eqs.~(\ref{eq:def-cond}) and (\ref
{eq:sigma-def}) yields 
\begin{equation}
G=\sigma (x_0,x_0^{\prime }),  \label{eq:2term-cond}
\end{equation}
where $x_0,x_0^{\prime }$ are arbitrary points inside the wire. The location
 of these points can be chosen so as to simplify  the calculation of $\sigma
(x_0,x_0^{\prime })$.

Let us consider again the case of a clean wire with an inhomogeneous
el--el interaction $V(x,y)$. For a 1D electron liquid the
Hamiltonian of the problem can be expressed in terms of conjugated bosonic
operators $\phi (x)$ and $\tilde{\phi}(x)$. The operator $\phi (x)$ is
related to the electron density operator $\rho (x)$ as $-\frac 1{\sqrt{\pi }}
d\phi (x)/dx=\rho (x)$; the operator $\tilde{\phi}(x)$ has a similar
relation with the current operator. The fluctuations of the charge density
are described by the Tomonaga--Luttinger Hamiltonian 
\begin{mathletters}
\begin{equation}
H_0=\frac{v_F}{2L}\sum_pp^2\tilde{\phi}_p\tilde{\phi}_{-p}+\frac{v_F}{2L}
\sum_{p,q}\left( \delta _{p,q}p^2+pqV(p,-q)/(v_F\pi L)\right) \phi _p\phi
_{-q},  \label{eq:Ham}
\end{equation}
where $\phi _p$ and $\tilde{\phi}_p$ are the Fourier transforms of the
operators $\phi (x)$ and $\tilde{\phi}(x)$, and $V(p,q)$ is the Fourier
transform of $V(x,y)$; here we set $e=\hbar =1$. When an external electric
field $E^{ext}(x,t)$ is applied, the term 
\begin{equation}
H_1=-\frac 1{\sqrt{\pi }L}\sum_p\phi _pE_p^{ext}(t)
\label{eq:external--field--action}
\end{equation}
should be added to the Hamiltonian. This term describes the interaction of
the local dipole moment with the external electric field.

The current operator in a 1D system is $J(p)=\frac i{\sqrt{\pi }}
\left[ H_0,\phi _p\right] =i\frac{v_Fp}{\sqrt{\pi }}\tilde{\phi}_p$, where
the commutation relations $\left[ \tilde{\phi}_{-q},\phi _p\right] =\frac Lp
\delta _{p,q}$ have been used. Then, the current $I=\left\langle
J\right\rangle $ induced by the external electric field is 
\end{mathletters}
\begin{equation}
I_{-\omega }(-q)=-i\frac{v_Fq}{\sqrt{\pi }}\left\langle \tilde{\phi}
_{-\omega ,-q}\right\rangle ,  \label{eq:current}
\end{equation}
where ${\displaystyle {\tilde{\phi}_{\omega ,q}=\int dte^{i\omega
t}\left\langle e^{i\left( H_0+H_1\right) t}\widetilde{\phi }(q)e^{-i\left(
H_0+H_1\right) t}\right\rangle }}$. As a result (see e.g., chapter 3 of 
Ref.~\cite{RFS:Mahan90}): 
\begin{equation}
I_{-\omega }(-q)=-i\frac{v_Fq}\pi \sum_pC_\omega (q,p)E_{\omega ,p}^{ext},
\label{eq:current1}
\end{equation}
where $C$ is the retarded correlation function of $\tilde{\phi}$ and $\phi $.
 The function $C$ obeys the Dyson equation 
\begin{equation}
C_\omega (q,p)=C_\omega ^0(p)\delta _{p,q}+C_\omega ^0(q)\frac 1{\pi L}
\sum_kqkV(q,-k)D_\omega (k,p),  \label{eq:K}
\end{equation}
where $D$ is the full propagator of $\phi $ and  $C_\omega ^0(p)=-\left(
1/2p\right) \left[ (\omega +v_Fp + i\gamma )^{-1} \right. + \left. (\omega
-v_Fp+i\gamma )^{-1}\right] $. The total electric field is the sum of the
external and the induced fields, $E^{tot}=E^{ext}+E^{ind}$. The
induced field $E^{ind}$ arises as a result of the redistribution of the
density of the electrons 
\begin{equation}
E_{\omega ,p}^{ind}=-\frac 1L\sqrt{\pi }\sum_qpqV(p,-q)\left\langle \phi
_{-\omega ,-q}\right\rangle .  \label{eq:Einda}
\end{equation}
Since $\left\langle \phi _{-\omega ,-q}\right\rangle =-\frac{v_F}{\sqrt{\pi }
}\sum_pD_\omega (q,p)E_{\omega ,p}^{ext}$, the induced field is related to $
E_{\omega ,p}^{ext}$ as 
\begin{equation}
E_{\omega ,p}^{ind}=\frac{v_F}L\sum_{qk}pqV(p,-q)D_\omega (q,k)E_{\omega
,k}^{ext}.  \label{eq:Eind}
\end{equation}
With the help of the Dyson equation~(\ref{eq:K}) and Eq.~(\ref{eq:Eind}) the
relation between $E_{\omega ,p}^{tot}$ and $E_{\omega ,q}^{ext}$ can be
obtained 
\begin{equation}
C_\omega ^0(p)E_{\omega ,p}^{tot}=\sum_qC_\omega (p,q)E_{\omega ,q}^{ext}.
\label{eq:Eext-Etot}
\end{equation}
This result corresponds to a well known fact in the diagrammatic technique,
that when the conductance is calculated with the help of the density
correlation function only the irreducible part of the correlation
function is involved. The response to the external electric field is
given by a series of diagrams containing polarization bubbles and starting
with an external field. The total field is given by diagrams of the same
type. Therefore the response to the total electric field is given by the
irreducible part of the correlation function. The importance of this fact
to the calculation of the conductance of quantum wires was emphasized
recently by Kawabata \cite{1DT:Kawabata96}. Substitution of Eq.~(\ref
{eq:Eext-Etot}) in the expression for the current, Eq.~(\ref{eq:current1}),
yields 
\begin{equation}
I_{-\omega }(-q)=-i\frac{qv_F}\pi C_\omega ^0(q)E_{\omega ,q}^{tot}
\label{eq:sigmapq}
\end{equation}
From this relation one can obtain the conductance $G$ for the two terminal
DC transport: 
\begin{equation}
G=\sigma (0,0)=\left( -i\right) \frac 1{\pi L}\sum_qqv_FC_{\omega
=0}^0\,(q)=1/2\pi .  \label{eq:sigma}
\end{equation}
Since the wire is attached to the reservoirs the electron states in the wire
have a finite width $\gamma ,$ which we assume to be larger than the level
spacing. Under this assumption the sum over momenta in Eq.~(\ref{eq:sigma})
was transformed to an integral. After restoring the constants $e$ and $\hbar 
$ the DC conductance of a clean wire becomes $G=e^2/2\pi \hbar $, i.e.,\
it is not influenced by the el--el interaction.

The above consideration was performed for an arbitrary el--el
interaction including the case when it is spatially inhomogeneous. Let us
discuss now a system of the type considered in Refs.~\cite{1DT:Maslov} in which the
el--el interaction exists only in the central part (of length $L_{int}$)
 and is absent in the segments attached to the central part of the
wire. One can check that if the DC conductance is calculated ignoring the
renormalization of the electric field, i.e., using $C_\omega (p,q) $ rather than 
 $C_\omega ^0(p) $ in Eq.~(\ref{eq:sigma}),  then there appear corrections 
 $\sim \left( \gamma L_{int}/v_F \right) \left(V/v_F \right) $.
 However, these corrections are not noticeable
when the region of the interaction, $L_{int}$, is short . In the treatment
of Refs.~\cite{1DT:Maslov} $v_F/\gamma $ corresponds to the length of the
wire $L$, and the limit $L_{int}/L\rightarrow 0$ was considered.

Let us discuss now a system with a backward scattering defect inside the
wire. Our goal now is to determine the effect of the renormalization of the 
external electric field on the conductance of this system.  We will follow the same line of
consideration as above. The conductance is given by 
\begin{equation}
G=\left( -i\right) \frac{v_F}{\pi L}\sum_{q,p}q\widetilde{{\cal C}}_{\omega
=0}^0\,(q,p),  \label{eq:gen--cona}
\end{equation}
where $\widetilde{{\cal C}}_{\omega =0}^0\,(q,p)$ is the irreducible (with
respect to the el--el interaction) part of the retarded
correlation function of the operators $\tilde{\phi}$ and $\phi $ in the
presence of the impurity backward scattering and the interaction. The full
correlation function $\widetilde{{\cal C}}\,$ is related to its irreducible
part via the Dyson equation 
\begin{equation}
\widetilde{{\cal C}}=\widetilde{{\cal C}}^0+\widetilde{{\cal C}}^0W
\widetilde{{\cal D}}.  \label{eq:Kt}
\end{equation}
Here $W(k,-q)= \frac 1{\pi L}qkV(q,-k) $ and the matrix $\widetilde{{\cal D}}$ is the
correlation function of $\phi $ operators. (Henceforth we use matrix notation.)
 The matrices $\widetilde{{\cal D}}
$ and $\widetilde{{\cal C}}$ carry information about the backward scattering
in the presence of interaction: 
\begin{equation}
\widetilde{{\cal D}}=D+D{\cal T}D,\;\widetilde{{\cal C}}=C+C{\cal T}D,
\label{eq:DKt}
\end{equation}
where ${\cal T}$ is the effective scattering matrix of the $\phi $--operators
 due to the impurity term, and the matrices $D$ and $C$ are the
correlators in the absence of the impurity.  After some transformations we obtain 
\begin{equation}
\widetilde{{\cal C}}^0=C^0\left( D^0\right) ^{-1}\left( \widetilde{{\cal D}}
^{-1}+W\right) ^{-1},  \label{eq:K0}
\end{equation}
where $D^0$ and $C^0$ are the irreducible parts of the correlators $D$ and $
C $, see Eq.~(\ref{eq:K}). Thus, the calculation of the conductance is
reduced to the inversion of  operators. To perform the inversion we will
assume that the impurity backward scattering is local, while the
el--el interaction inside the wire is homogeneous and short
range, i.e., the elements of the matrix ${\cal T}$ do not depend on the
momenta and $W(q,-k)=\frac 1\pi V_0q^2\delta _{q,k}$. Now 
the inversion can be done straightforwardly and one obtains 
\begin{equation}
\widetilde{{\cal C}}^0=C^0+\frac 1{1+\text{{\it tr}}{\cal T}\left(
D-D^0\right) }C^0{\cal T}D^0.  \label{eq:K01}
\end{equation}
With the use of Eq.~(\ref{eq:gen--cona}) the conductance $G\left( {\cal T}
\right) $ is determined if the scattering matrix ${\cal T}$ is known.

 When one ignores the effect of the renormalization of the electric field
 the full correlator $\widetilde{{\cal C}}$ is used 
instead of $\widetilde{{\cal C}}^0$ in Eq.~(\ref{eq:gen--cona}).
 The quantity obtained will be denoted by $G^{\prime }$. Contrary to the 
conductance $G $, which is the response to the total electric field, 
 $G^{\prime} $ describes the response to the external field. 
Using Eq.~(\ref{eq:DKt}) $G^{\prime } \left( {\cal T}\right) $ can be found 
\begin{equation}
G^{\prime }\left( {\cal T}\right) =\left( -i\right) \frac{v_F}{\pi L}
\sum_{q,p}qC_{\omega =0}(q)\left( \delta _{q,p}+{\cal T}(q,p)D_{\omega
=0}(p)\right) \,.  \label{eq:gen--conb}
\end{equation}
The relations (\ref{eq:K01}) and (\ref{eq:gen--conb}) enable us to exclude $
{\cal T}$ and to express $G$ as a function of $G^{\prime }$ 
\begin{equation}
G=\frac{K_\rho G^{\prime }}{2\pi (K_\rho -1)G^{\prime }+K_\rho },
\label{eq:G1G}
\end{equation}
where $K_\rho =1\left/ \sqrt{1+V_0/v_F\pi }\right. $.  For a clean wire 
$G^{\prime }=K_\rho /2\pi $ and Eq.~(\ref{eq:G1G}) reproduces the universal
value of the conductance $G$. The structure of Eq.~(\ref{eq:G1G}) reflects
the fact that in the presence of a backward scattering center the effect of
the electric field renormalization depends not only on the el--el
interaction, but also on the interplay between the backscattering and the
el--el interaction. The quantity $G^{\prime }$, rather than $G$, has been
extensively studied in the recent years by diverse techniques \cite
{1DT:Kane92aFurusaki93Ogata94,EE1D:Fendley95,1DT:Weiss96} in the
perturbative and nonperturbative regimes. Eq.~(\ref{eq:G1G}) allows one to
use these results to recalculate the conductance in order to include the
effect of the renormalization of the electric field.

Until now the simplified case of a single mode wire has been considered. In
a real quantum wire a few modes exist due to spin and subbands corresponding
to quantization of transversal motion. In the absence of backward scattering
Eq.~(\ref{eq:sigma}) can be easily generalized to the case when $N>1$ modes
are occupied. Namely, $v_FC_{\omega =0}^0\,(q)$ should be substituted by $
\sum_{n=1}^Nv_F^nC_{\omega =0}^{0,n}(q)$ where $n$ is the mode index. This
yields the conductance of a multimode wire $G_N=Ne^2/2\pi \hbar $. To
generalize Eq.~(\ref{eq:G1G}) we consider the most symmetric case when the
Fermi velocities in all channels are identical, $v_F^n=v_F$, and only 
an interaction of the form $\rho(x)V_0\delta (x-y)\rho (y)$ is present, 
where $\rho $ is the total electron density. Then 
\begin{equation}
G_N=\frac{K_\rho ^NG_N^{\prime }}{2\pi (\frac{K_\rho ^N-1}N)G_N^{\prime
}+K_\rho ^N},  \label{eq:G1GN}
\end{equation}
where $K_\rho ^N=1\left/ \sqrt{1+NV_0/v_F\pi }\right. $ and the
conductance--like quantity $G_N^{\prime }$ is the response to the external
electric field. In the absence of backward scattering $G_N^{\prime }=NK_\rho
^N/2\pi $, resulting in $G_N=Ne^2/2\pi \hbar $.

To summarize, we have studied the influence of the el--el
interaction on the two terminal conductance of quantum wires. It was
shown, by two different approaches, that the universal value of the
conductance in a clean wire is a result of a cancelation of the
effects of the el--el interaction on the current and on
the external electric field. In addition for a system with a backward
scattering center we have found the relation of the DC conductance to
 the response to the external electric field.

The last remark concerns the relation of the edge state electrons,
under the condition of the quantum Hall effect (QHE), to the
interacting 1D electron gas \cite{FQHE:Wen91a}. It is
a rather common believe that the physics of the edge states in the
fractional QHE with $\nu =1/\left( 2p+1\right) $ and the physics of
the interacting 1D electron gas are equivalent when the filling factor
$\nu =K_\rho $. The fact that $G^{\prime }=K_\rho e^2/2\pi \hbar $ and
the Hall conductance $\sigma _{xy}=\nu e^2/2\pi \hbar $ is one of the
reasons for that point of view. In this connection we would like to
emphasize that $\nu $ is not completely equivalent to $K_\rho $.  In
the fractional QHE the filling factor $\nu $ appears through the
commutation relations of the operators of the electron density, but
not as a result of the density--density interaction of the edge state
electrons. For that reason the effect of the electric field
renormalization has no connection with the factor $\nu $, and therefore 
 $\sigma _{xy}$ does contain it. On the other hand, the Hall
conductance of the edge states is not modified by an interedge
el--el interaction \cite {QHE:Oreg95}, precisely in the
same way as in the case of a clean wire.

We thank A.~Kamenev, D.~Orgad and A.~Stern for useful discussions. A.~F. is
grateful to H.~Fukuyama, A.~Furusaki, N.~Nagaosa, M.~Ogata and S.~Tarucha
for a few highly stimulating discussions, and for the hospitality during his
visit at the University of Tokyo. A.~F. is acknowledged to the JSPS for this
visit. A.~F. is supported by the Barecha Fund Award. This work is supported
by the Israel Academy of Science, Grant No. 801/94-1 and by the
German--Israel Foundation (GIF).

\bibliographystyle{prsty}

\begin{thebibliography}{10}

\bibitem[\dag]{Lan}
Also at the Landau Institute for Theoretical Physics, Russia.

\bibitem{1DT:Landauer}
{R}. Landauer, {P}hil. Mag. {\bf 21}, 863 (1970); {Y}. Imry, in {\em Direction
  in condence matter physics}, edited by G. Grinstain and G. Mazenko (World
  Scientific Publishing Co., Singapore, 1986), pp.\ 101 and references therin.

\bibitem{1DT:Apel82}
W. Apel and T.~M. Rice, Phys. Rev. B {\bf 26},  7063  (1982).

\bibitem{QHE:Oreg95}
Y. Oreg and A.~M. Finkel'stein, Phys. Rev. Lett. {\bf 74},  3668  (1995).

\bibitem{1DT:Kawabata96}
A. Kawabata, J.~ Phys. Soc. Jap. {\bf 65},  30   (1996).

\bibitem{1DT:Tarucha95}
S. Tarucha, T. Honda, and T. Saku, Sol. Stat. Com. {\bf 94},  413   (1995).

\bibitem{1DT:Kane92aFurusaki93Ogata94}
{C}.~L. Kane and M.~P.~A. Fisher, Phys. Rev. B {\bf 46}, 15233 (1992); {A}.
  Furusaki and N. Nagaosa, Phys. Rev. B {\bf 47}, 4631 (1993); {M}. Ogata and
  H. Fukuyama, Phys. Rev. Lett. {\bf 73}, 468 (1994).

\bibitem{1DT:Maslov}
{D}.~L.~Maslov and M.~Stone, Phys. Rev. B {\bf 52} R5539 (1995); {V}.~V.
  Ponomarenko, Phys. Rev. B {\bf 52}, R8666 (1995); I. Safi and H. Schulz,
  Phys. Rev. B {\bf 52}, R17040 (1995).

\bibitem{1DT:Kane88}
C.~L. Kane, R.~A. Serota, and P.~A. Lee, Phys. Rev. B {\bf 37},  6701   (1988).

\bibitem{EE1D:Mattis65}
D.~C. Mattis and E.~H. Lieb, J. Math. Phys. {\bf 6},  304  (1965).

\bibitem{RFS:Mahan90}
G.~D. Mahan, {\em Many-Particle Physics, Second Edition} (Plenum Press,
  New-York and London, 1990).

\bibitem{EE1D:Fendley95}
P. Fendley, A.~W.~W. Ludwig, and H. Saluer, Phys. Rev. Lett. {\bf 74},  3005
  (1995).

\bibitem{1DT:Weiss96}
U. Weiss, cond--mat {\bf 04},  027  (1996).

\bibitem{FQHE:Wen91a}
X.~G. Wen, Phys. Rev. B {\bf 43},  11025  (1991).

\end{thebibliography}

\end{document}